\documentclass[twocolumn]{aastex62}
\usepackage{xcolor}
\usepackage{amsmath}
\usepackage{mathrsfs}
\usepackage{hyperref}

\newcommand\fitburst{\textsc{fitburst}}
\definecolor{darkblue}{rgb}{0.0, 0.0, 0.7}
\graphicspath{{./}{figures/}}

\begin{document}
\shortauthors{Fan et al.}

\author[0000-0002-0786-7307]{Bo Lin Fan}%
\affiliation{Dunlap Institute for Astronomy and Astrophysics, University of Toronto, Toronto, Canada}
\affiliation{David A. Dunlap Department for Astronomy and Astrophysics, University of Toronto, 50 St George Street, Toronto, Canada ON M5S 3H4}
\author[0000-0002-0965-7864]{Ren\'ee Hlo\v{z}ek}%
\affiliation{Dunlap Institute for Astronomy and Astrophysics, University of Toronto, Toronto, Canada}
\affiliation{David A. Dunlap Department for Astronomy and Astrophysics, University of Toronto, 50 St George Street, Toronto, Canada ON M5S 3H4}
\author[0000-0002-3654-4662]{Antonio Herrera Martin}
\affiliation{Department of Statistical Sciences, University of Toronto,  700 University Ave 9th Floor, Toronto, Canada, ON M5G 1X6}
\affiliation{David A. Dunlap Department for Astronomy and Astrophysics, University of Toronto, 50 St George Street, Toronto, Canada ON M5S 3H4}

\title{Semi-supervised morphological classification of fast radio bursts from the second CHIME/FRB catalogue}
\begin{abstract}
{Understanding the morphology of fast radio bursts (FRB), and whether all sources repeat, are key challenges that are becoming more tractable given the increase in data from surveys such as the Canadian Hydrogen Intensity Mapping Experiment FRB project (CHIME/FRB). We present a Convolutional Autoencoder unsupervised classifier for separating the CHIME/FRB data into morphological classes. This data-driven approach is more reproducible than visual inspection, since groupings are learned from the data itself and not subject to differences between expert annotations. While most bursts occupy a similar area of morphological parameter space, we identify three classes of bursts separate from the general FRB population. While one class contains bursts with short bandwidth, and downward-drifting sub-burst structure, the characteristic bursts of other classes have very short temporal width, and occupy the entire CHIME observing band.  We identify two distinct subgroups of temporally short, simple broadband bursts; one with minimal scattering and the other with higher scattering. As an additional output of our classifier, we provide a binary FRB repeatability classification, and train the classifications on simulations that mimic the first FRB catalogue from CHIME/FRB. We are able to correctly identify 86$\%$ of repeater bursts. We find that our approach is able to independently recover the downward linear drifting burst morphologies previously defined through visually inspection. Overall, we find that although there exists FRB subgroups with higher or lower proportion of repeaters, there is substantial overlap between the morphological properties of repeaters and one-off bursts consistent with previous studies.}
\end{abstract}


\section{Introduction} \label{sec:intro}
Since their discovery \citep{lorimer/etal:2007}, the classification of  fast radio bursts (FRBs) based on their spectral morphology has become an active area of research. Understanding the morphological differences between groups of FRBs may yield insight into the physical origins or progenitor systems of FRBs. The properties of these luminous, microsecond-millisecond timescale radio transients are mostly consistent with an extragalactic origin \citep{petroff/etal:2022}, although the FRB-like object FRB 20200428 was localized to our galaxy and associated with a Galactic magnetar SGR 1935+2154 \citep{bochenek/etal:2020,chime2020galactic_magnetar}. The precise physical mechanisms responsible for FRB emission remain elusive \cite{wang/etal:2022,geng/etal:2020}. While early FRB detections were made using telescopes such as Murriyang, the Parkes radio telescope \citep{parks/2016}, the field has advanced rapidly with the recent advent of new facilities, including the Deep Synoptic Array-110 \citep[DSA-110;][]{2026DSA-110}, the Commensal Real-time ASKAP Fast Transient \citep[CRAFT;][]{2025craft} Survey, the  MeerKAT radio telescope \citep[the MeerTRAP sample;][]{2022meertrap}, Five-hundred-meter Aperture Spherical radio Telescope \citep[FAST;][]{fast/2011}) and the Canadian Hydrogen Intensity Mapping Experiment \citep[CHIME;][]{chime/2022}. FAST and CHIME/FRB in particular have increased the number of FRBs by orders of magnitude in the short time they have been operating, with the former recording $\sim 10,000$ bursts from a few actively repeating sources, and the latter containing from 3641 unique sources \citep{fast/2011,dr2}. The detection of bursts from a large number of unique sources makes the second CHIME/FRB catalogue useful for enabling population-level studies of FRBs \citep{dr2}. While the FRB emission mechanism is not yet known, several possible explanations have been proposed, including their originating from compact object outflows that cause relativistic shocks in surrounding plasma, or interactions in a neutron star magnetosphere \citep{2017kumar,2019metzger,zhang/etal:2020,petroff/etal:2022}. Excitingly, rather than being detected as a one-off burst, some FRBs have been found to repeat, suggesting that either all FRBs are generated through some repeatable mechanisms, or that there are two or more populations of FRBs, each with their own emission mechanism. Only about ~3$\%$ of detected FRBs have been shown to repeat \citep{cook2026discovery30repeatingfast}. We will hereby refer to the property of whether a FRB originates from a source that has been known to repeat as its `repeatability.'

With the advent of the first CHIME/FRB catalogue (DR1) \citep{chime/2022}, population-level studies can be conducted to determine possible groupings in the morphological characteristics of FRBs, as well as correlations between FRB morphological characteristics and repetition. Particularly, CHIME/FRB uses the \fitburst\ pulse modelling algorithm \citep{fonseca2024fitburst} to fit analytical pulse shapes to the dedispersed frequency-time data (waterfall plot) of FRBs. The FRBs are then categorized based on morphological parameters such as bandwidth and temporal width obtained by \fitburst. A generally agreed upon morphological trend is that repeaters have been found to have smaller spectral extent and higher temporal width than one-off pulses \citep{pleunis/etal:2021,curtin/etal:2025}. First found in \citet{pleunis/etal:2021}, this relation has been demonstrated to be robust in followup studies such as \citet{chen2021chime,sharma2024}. However, studies aiming to predict which FRBs repeat based on morphological parameters tend to overestimate the repeater fraction, with $> 50 \%$ of predicted repeaters originating from non-repeating sources \citep{chen2021chime,sun2026dr1_analysis,sun2026dr2_analysis,kumar/etal:2025}.

With the release of the second CHIME/FRB catalogue (DR2), further investigation showed that side-lobe effects may explain some fraction of these one-off events that were misidentified as repeaters \citep{dr2}, as bursts detected in CHIME instrument's side lobe may have part of their spectrum suppressed, leading them to appear more narrow-band, and therefore to seem similar to the repeater morphology described in \citet{pleunis/etal:2021}. 

In addition to side-lobe effects, the beam-forming software used in CHIME/FRB intensity data leads to underestimation of burst bandwidth, as baseband observations of the same bursts reveal higher estimated bandwidths \citep{sand/etal:2024/morphology,bridget2023beam}. Other works have found similar repeatability trends in DR2 as compared to DR1. For example, \citet{sun2026dr2_analysis} uses Uniform Manifold Approximation and Projection \citep[UMAP;][]{umap} dimensionality reduction, followed by clustering of the \fitburst\ parameters of the DR2 bursts, similar to the methodology applied to DR1 in \citet{sun2026dr1_analysis}. They find that repeaters in DR2 exhibit more spectrally narrow morphology than one-off bursts, with the exception of a few repeater bursts with one-off-like, broadband, temporally narrow characteristics. Analyzing burst rates, \citet{cook2026discovery30repeatingfast} finds that it cannot be ruled out that one-off bursts and repeaters may be drawn from the same population. Furthermore, \citet{cook2026discovery30repeatingfast} suggests that large FRB surveys such as CHIME/FRB are biased \textit{against} detecting repeater-like signals, indicating that FRBs that repeat on long timescales may be mischaracterized as one-off events. Overall, it is not known what proportion of these repeater-like non-repeaters are caused by morphological overlap, instrumental effects, or originate from intrinsically repeating FRBs.

Beyond analyzing the parameters obtained through the \fitburst\ modelling procedure, analyzing the frequency-time waterfall plots themselves can give additional information on burst morphology, such as a potential sub-burst structure in FRBs. As shown in \citet{kharel/etal:2025}, for the CHIME DR2 repeaters presented in \cite{dr2}, morphology analysis applied to the full waterfall plot data alone goes some way to distinguishing between repeating and non-repeating FRBs. As a result, there has been an increased effort to identify morphological groups of FRBs based on analyzing their waterfall plots. Burst morphologies are typically characterized into groupings visually with machine learning methods using these visually-defined groups as training labels \citep{kumar/etal:2025}, or an unlabelled training dataset for deep learning burst classification \citep{kuiper/etal:2025}. It remains to be shown whether these visually defined morphological categories can be recreated through waterfall plot data alone.

In this work we train a Convolutional Autoencoder \citep[CAE;][]{cae} unsupervised classifier to separate bursts in the CHIME/FRB into morphological classes, and train (as an additional output) a binary repeatability classification. 
CAEs are suited for this task since convolution can capture local spatial (in our case, in frequency-time space) correlations in FRB waterfall plots. Since our goal is to discover morphological subtypes rather than reconstruction accuracy, we use a CAE for its ability rather than related architectures such as variational auto-encoders (VAEs) since it does not enforce the learned latent parameters to follow a Gaussian distribution \citep{2024aes}. 

We discuss the CHIME/FRB dataset in \autoref{sec:data} and our procedure for simulating a training set of FRBs in \autoref{sec:simulation}. In contrast to previous approaches where visually defined classes were used as training data, we apply representation learning directly to real CHIME/FRB intensity data to determine whether morphological classes emerge naturally, without the need for manually defining classes, and whether they correspond to repeatability. We describe the CAE used to create a lower dimensional representation of CHIME/FRB waterfall plot data in \autoref{sec:cae}. Unsupervised analysis of lower dimensional latent representation is used to define categories of bursts, while a classification network is simultaneously trained on this latent representation to predict repeatability. We present the results in \autoref{sec:results}. We are able to recover previously defined morphologies, justifying their classification with data, and identify a few previously not defined groups of bursts, showing that unsupervised categorization can give new insight into structure beyond visual identification. Additionally, we also seek to determine whether repeatability can be linked to these morphological groups. While repeaters and one-off bursts do not form completely separable populations in latent space, repeater-rich morphological classes exhibit statistically distinct spectral and temporal properties consistent with previous studies, as we show in \autoref{sec:repeatability}. We discuss our results and potential next steps in \autoref{sec:end}.

\section{Data and Simulations}
We train an FRB simulation model on properties of the first data release from CHIME, and discuss both DR1 and DR2 and their observational properties.
\label{sec:data}
\begin{figure}[htbp!]
\begin{center}
\begin{tabular}{c}
\includegraphics[width=0.45\textwidth]{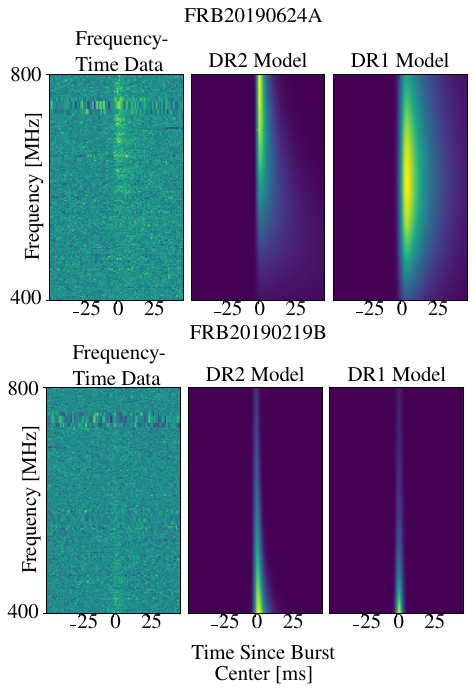}
\end{tabular}
\end{center}
\caption{Examples of bursts modelled using the \fitburst~implementation in both the second \textit{(middle panel)} and first \textit{(right panel)} CHIME/FRB data release papers, compared to the CHIME/FRB total intensity image \textit{(left panel)}. The top row shows the data and model for FRB 20190219B, which shows some changes to the model parameters between data releases, while the bottom row is for FRB 20181213A, a burst that is consistent across releases. We note that the updated \fitburst\ model has not been released and as such the middle and right panels in each row are produced by using the updated \fitburst\ parameters in the \citet{fonseca2024fitburst}. While the burst modelling is largely consistent models are similar between data-releases, the modelling of some bursts has changed between releases, as described in \citet{dr2}. We use \fitburst\ to simulate FRBs for the classification training, as described in \autoref{sec:simulation}.\label{fig:fitburst_examples} }
\end{figure}
\subsection{CHIME/FRB Intensity catalogue}
\label{sec:data_cat}
The CHIME telescope is located at the Dominion Radio Astrophysical Observatory (DRAO), near Penticton, British Columbia, and operates over a frequency range of 400–800 MHz \citep{chime_orig}. CHIME/FRB intensity data has a temporal resolution of 0.983ms. The second CHIME/FRB catalogue provides the largest sample of FRBs to date \citep{dr2}. 

Morphological characteristics of bursts presented in CHIME/FRB data are estimated using the \fitburst\ algorithm \citep{fonseca2024fitburst}. This algorithm models the de-dispersed time-frequency structure (i.e. a so-called `waterfall plot') of each event using analytic pulse shapes, such as Gaussian or scatter-broadened Gaussian components. This modelling enables pulse properties including width, scattering timescale, spectral index, and spectral running to be estimated by \fitburst. For multi-component bursts, individual sub-bursts are fit together in order to recover the full temporal extent and spectral structure of the event. These uniformly-derived parameters provide a standardized basis for linking parametric burst properties to morphological categorizations we present in this paper.

Two CHIME/FRB catalogues have been released to date: Data Release 1 \citep[DR1,][]{chime/2022} includes 536 bursts, 98 of which originate from 18 unique, repeating FRBs recorded between 2018 July 25 to 2019 July 1 while Data Release 2 \citep[DR2,][]{dr2} contains 4539 pulses, including 981 bursts from 83 known repeating sources. Re-analyzed events taken from DR1 are included in DR2 data as described in \citet{dr2}. The CHIME/FRB DR1 and DR2 datasets were obtained using the same instrument and observational setup, and therefore the methodology for obtaining dedispersed waterfall plots is unchanged between the two releases. The primary differences between DR1 and DR2 arise from updates to the \fitburst\ modelling pipeline used to detect and fit the FRB burst profiles. As stated in \citet{dr2}, improvements were made to the numerical implementation of the model with the DR2 release, including a more stable treatment of scattered pulse profiles and the removal of manually-determined parameter prior bounds previously used when fitting multi-component bursts. We show an example of two bursts included in both data releases in \autoref{fig:fitburst_examples}. Despite these methodological updates, the majority of bursts are modelled consistently across \fitburst~implementation used in the two catalogues, with approximately 78$\%$ of DR1 bursts re-fitted in DR2 having unchanged \fitburst~models. We will describe later, however, how the changes impact clustering on morphological parameters.
\begin{figure*}[htbp!]
\begin{center}
\begin{tabular}{c}
\includegraphics[width=0.95\textwidth]{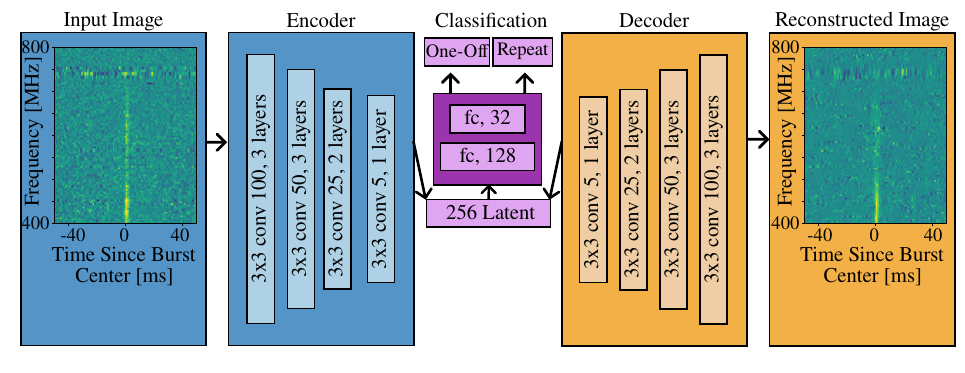}
\end{tabular}
\end{center}
\caption{The flow diagram of the semi-supervised CAE used to determine FRB morphology while simultaneously classifying for FRB repeatability. A waterfall plot of a given burst \textit{(left)} is encoded into a lower dimensional representation, the latent space. The model performance is then evaluated as a sum of its ability to reconstruct the original burst waterfall plot from this latent representation, and its ability to accurately classify a burst as repeating or one-off from this latent parametrization \textit{(middle)}. The reconstructed waterfall plot is also provided as an output of the procedure \textit{(right)}. \label{fig:CAE} }
\end{figure*}

\subsection{Simulating FRB Observations using \fitburst}
\label{sec:simulation}
The CHIME/FRB datasets are both limited in size and unbalanced with respect to burst type. The CHIME/FRB DR1 catalogue contains of 536 bursts, and in both DR1 and DR2 only approximately one fifth of events originate from known repeating sources. This imbalance between repeating and non-repeating events presents a challenge for representation learning, as CAEs trained directly on the observed data may preferentially learn features associated with the single/one-off bursts, reducing their ability to generalize to repeaters or rarer morphologies \citep{huang2016smallset}. Such issues are well documented in machine learning applications involving imbalanced datasets, where minority classes are often underrepresented in the learned feature space \citep[see e.g.][]{huang2016smallset}. To address these limitations, we generate a synthetic training set designed to capture the diversity of observed FRB properties while ensuring balanced representation of burst types. Importantly, only DR1 bursts are used to construct these simulations. This preserves DR2 as a largely independent dataset for further analysis and enables us to test whether morphological structure identified from DR1-informed training generalizes to the substantially larger DR2 sample. We note that while these \fitburst~parameters are biased by CHIME's beam effects, they offer a way to reproduce the observed distribution of burst properties and generate an empirically motivated set of training examples that reflect the diversity of bursts present in real data.

To construct realistic population-level simulations, we fit multi-dimensional kernel density estimates (KDEs) to the observed \fitburst\ parameter distributions of one-off and repeating FRBs in DR1 separately. The KDEs are defined over temporal width, spectral index, spectral running, and scattering timescale. Synthetic burst parameters are then drawn from these distributions to generate simulated events representative of each population. For many bursts in the CHIME/FRB catalogue, central frequency ($\nu_c$) lies at the edges of the CHIME/FRB observational bandwidth ($\nu_c < 400$ MHz or $\nu_c > 800$ MHz for lower and upper bounds respectively). This implies that the true central frequency for these bursts lie outside of the CHIME/FRB observing band \citep{pleunis/etal:2021}. Since the true distribution of CHIME/FRB DR1 $\nu_c$ is unknown, we aim to sample from bursts above or below the observing band by drawing $\nu_c$ it from a uniform distribution between 200 and 1200 MHz. Similarly, the distribution of scattering indices is not known for the CHIME/FRB DR1 sample and may be degenerate with other mechanisms that cause frequency-dependent broadening in time, such as intrinsic frequency drifting on errors when estimating dispersion measure. As such, we sampled scattering index from a uniform distribution ranging from -4.4 to –3.5 in accordance with the empirical range determined by \citet{petroff2016cata}.

Using these distributions, we simulate 2500 one-off pulses for both the one-off and repeater groups, ensuring even coverage of both types of burst. In order to simulate bursts with multiple sub-bursts, we simulate additional sub-pulses with the same spectral index, spectral running, and temporal width as the initial pulse. The sample of FRBs containing multiple sub-bursts is small, and baseband observations often reveal additional sub-bursts that cannot be obtained using CHIME/FRB total intensity data temporal resolution \citep{faber/etal:2024}. \citet{faber/etal:2024} finds that bursts may contain up to 8 sub-bursts with drift rates up to 30MHz ms$^{-1}$ in CHIME/FRB baseband data, though the majority of these sub-bursts cannot be resolved in CHIME/FRB total intensity data. So, to simulate a diverse range of sub-burst structure, we generate between one and four sub-bursts are generated per event, with their central times, frequencies, and bandwidths sampled uniformly from random offsets ranging from 5 ms to 10 ms relative to the preceding pulse. Central frequency differences of each sub-burst and relative to the preceding burst were sampled from a random uniform distribution between -200 MHz and 10 MHz. This is to encompass both typical downward-drifting bursts and the rarer upward-drifting events reported in the literature \citep{sand/etal:2024/morphology,faber/etal:2024}. We choose to include a small number of upward drifting bursts to have a more balanced training set. We simulate 2500 multiple-sub-burst events sampling from both the one-off burst and repeater parameter distributions. Parameters presented in \autoref{table:fitburst_params} are applied in \fitburst\ as described in \citet{fonseca2024fitburst} to simulate a range of FRB morphologies. 

\begin{figure*}[htbp!]
\begin{center}
\begin{tabular}{c}
\includegraphics[width=0.75\textwidth]{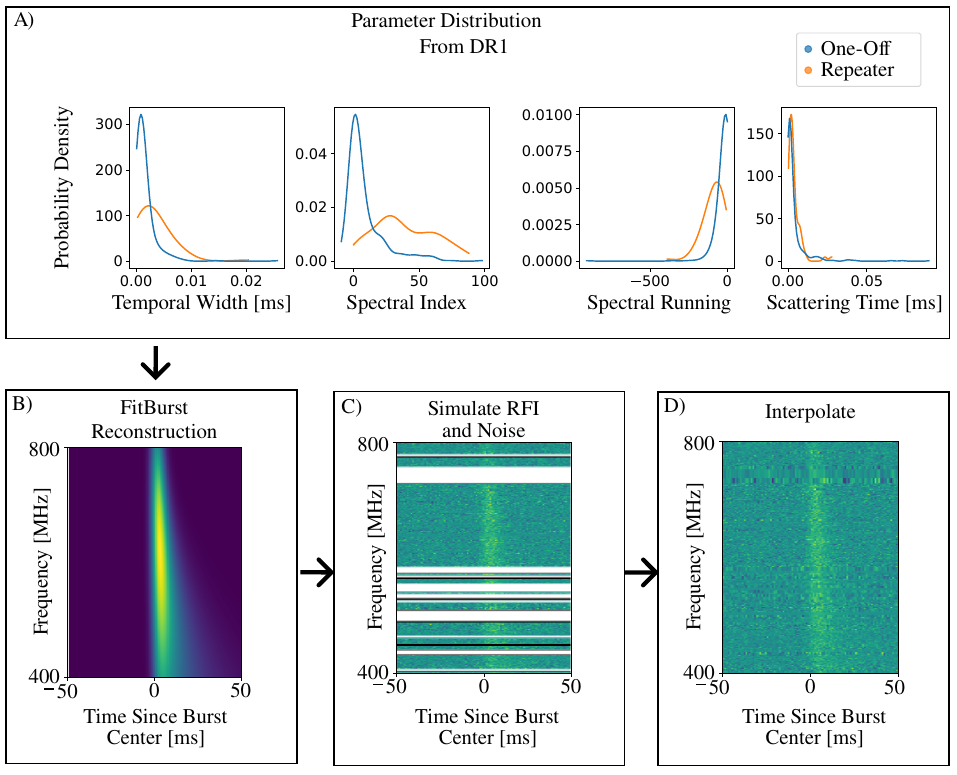}
\end{tabular}
\end{center}
\caption{In order to simulate bursts to train our CAE, we first sample parameters shown in the box A) from KDE fitted to repeating or non-repeating FRBs in the first CHIME/FRB dataset. These parameters are then used to simulated burst waterfall plots using \fitburst\ (B). Finally, Gaussian noise is applied and certain frequency bands are masked to simulate CHIME instrumental noise and RFI respectively (C), and the plots are interpolated with noise in the masked regions (D). \label{fig:simulation} }
\end{figure*}

Radio Frequency Interference (RFI) is the broad term for unwanted, human-made, radio emissions that can be identified as higher intensity in given frequency bands compared to bands without RFI \citep{rafiei2023rfi}. Both thermal noise and terrestrial RFI influence FRB waterfall plots, and can make it challenging to determine burst morphology. As discussed in \citet{chime/2022}, thermal and instrumental noise results in white noise across observing band, while RFI causes the loss of specific bands. To replicate observational noise, we incorporate RFI masking and instrumental noise into our simulations. We find the probability that each given frequency channel is masked due to RFI in CHIME/FRB DR1 and use that as the probability the frequency channel is masked in our simulated burst. Thermal noise is also simulated for each frequency channel by taking theroot-mean-square (RMS) noise distribution measured in the DR1 intensity data, and adding Gaussian noise with this RMS to our simulated waterfall plots. Our simulation pipeline is illustrated in \autoref{fig:simulation}.

\subsection{Denoising and Processing} \label{sec:denoise}
Prior to input into CAE, all observed CHIME/FRB bursts from both catalogues and simulated bursts described in \autoref{sec:simulation} were preprocessed using the same method. Bursts are down-sampled to 1024 frequency bands from 16384 bands using the publicly released preprocessing notebook\footnote{Obtained from \href{https://chime-frb-open-data.github.io/}{https://chime-frb-open-data.github.io/}.}. The RFI identified by the CHIME/FRB Collaboration are also removed using this notebook. Additional denoising of the data was performed in two stages. Firstly, we applied a 5 pixel by 5 pixel Gaussian smoothing filter to suppress white noise in the waterfall plot. Secondly, to mitigate residual RFI and particularly noisy frequency channels, we filter with a $100\times1$ pixel tophat filter along the frequency axis while ignoring RFI masked bands. This reduce channel-to-channel variations and produces a smoother estimate of the underlying burst spectrum for further resizing and interpolation steps. The resulting waterfall plots were then resized to a uniform resolution of 100 frequency bins by 100 time bins, with padding by white noise where necessary (so that any classification algorithm does not concentrate on parts of the image that are explicitly zero), in order to standardize resolution before inputting into our CAE.

\section{Methods} \label{sec:methods}

\subsection{Unsupervised Convolutional Autoencoders} \label{sec:cae}
Autoencoders have been used in FRB analyses to derive compact representations of \fitburst\ parameters \citep{arni2025usingdeeplearningrobust,mankatwit2025semisup} and to generate embeddings of visually defined morphological groups directly from waterfall plots \citep{kumar/etal:2025,kuiper/etal:2025}. Because CAEs operate on the full waterfall plot rather than a limited set of fitted quantities, they provide a flexible and minimally assumption-driven approach to learning burst morphology \citep{cae}.

A CAE consists of an encoder that compresses the input image into a low-dimensional latent representation and a decoder that reconstructs the image from this compressed encoding \citep{cae,Guo_CAE,zhang2018better}. The model is trained by minimizing the reconstruction loss, defined as the difference between the input and reconstructed image. In our implementation, the encoder comprises nine convolutional layers that progressively reduce the spatial dimensionality of the input, followed by a fully connected bottleneck layer containing 256 latent parameters. The decoder mirrors the encoder with nine convolutional layers that reconstruct the waterfall plot from this latent representation. All convolutional layers use Rectified Linear Unit \citep[ReLU,][]{relu} activation functions, and model optimization is performed using the Adaptive Moment estimation \citep[known as ``Adam'',][]{adam} optimizer. A schematic overview of the architecture is shown in \autoref{fig:CAE}, and a detailed layer-by-layer description with training hyper-parameters is provided in \autoref{table:unsupervised_params} of the Appendix.
\subsection{Supervised Classification ``Head''} \label{sec:supervised_layer}
In addition to reconstruction, we simultaneously train a supervised classification ``head'' using the 256-dimensional latent representations as input. This classifier is designed to predict whether a burst originates from a repeating or apparently single source. Simulated bursts drawn from the one-off parameter distribution are labelled as non-repeaters, while those drawn from the repeater distribution are labelled as repeaters. The classifier consists of three fully connected linear layers with 128, 32, and 2 neurons, respectively, and uses ReLU activation functions between hidden layers. The CAE is trained on 7,500 (3750 repeating, 3750 one-off) simulated bursts and validated on an independent set of 2,500 (1250 repeating, 1250 one-off) simulated bursts, as described in \autoref{sec:data}. These simulations provide equal coverage of repeating and one-off morphologies, enabling stable representation learning. We evaluate the performance of each algorithm by computing their values of Precision and Recall as:
\begin{eqnarray}
\mathrm{Precision}&=&{\frac {\mathrm{TP}}{\mathrm{TP}+\mathrm{FP}}} \nonumber \\
\mathrm{Recall}&=&{\frac {\mathrm{TP}}{\mathrm{TP}+\mathrm{FN}}}
\label{eq:precision_recall}
\end{eqnarray}

We calculate a number of True Positive (TP) repeater classifications as the number of observed repeaters predicted to be repeaters. False Positive (FP) and False negative (FN) events are defined as the number of apparent one-off bursts classified as repeaters, and the number of observed repeaters classified as one-off bursts respectively.

\subsection{Dimensionality reduction}
Dimensionality reduction is an effective approach for identifying structure in high-dimensional datasets and for grouping objects into distinct morphological classes \citep{raquel2024unsup,chen2023unsup,Luo_2022unsup,sharma2024}. We use dimensionality reduction techniques on the CAE embedding parameters in order to visualize how bursts of differing properties may cluster.
\begin{figure*} [htbp!]
\begin{center}
\begin{tabular}{c}
\includegraphics[width=0.7\textwidth]{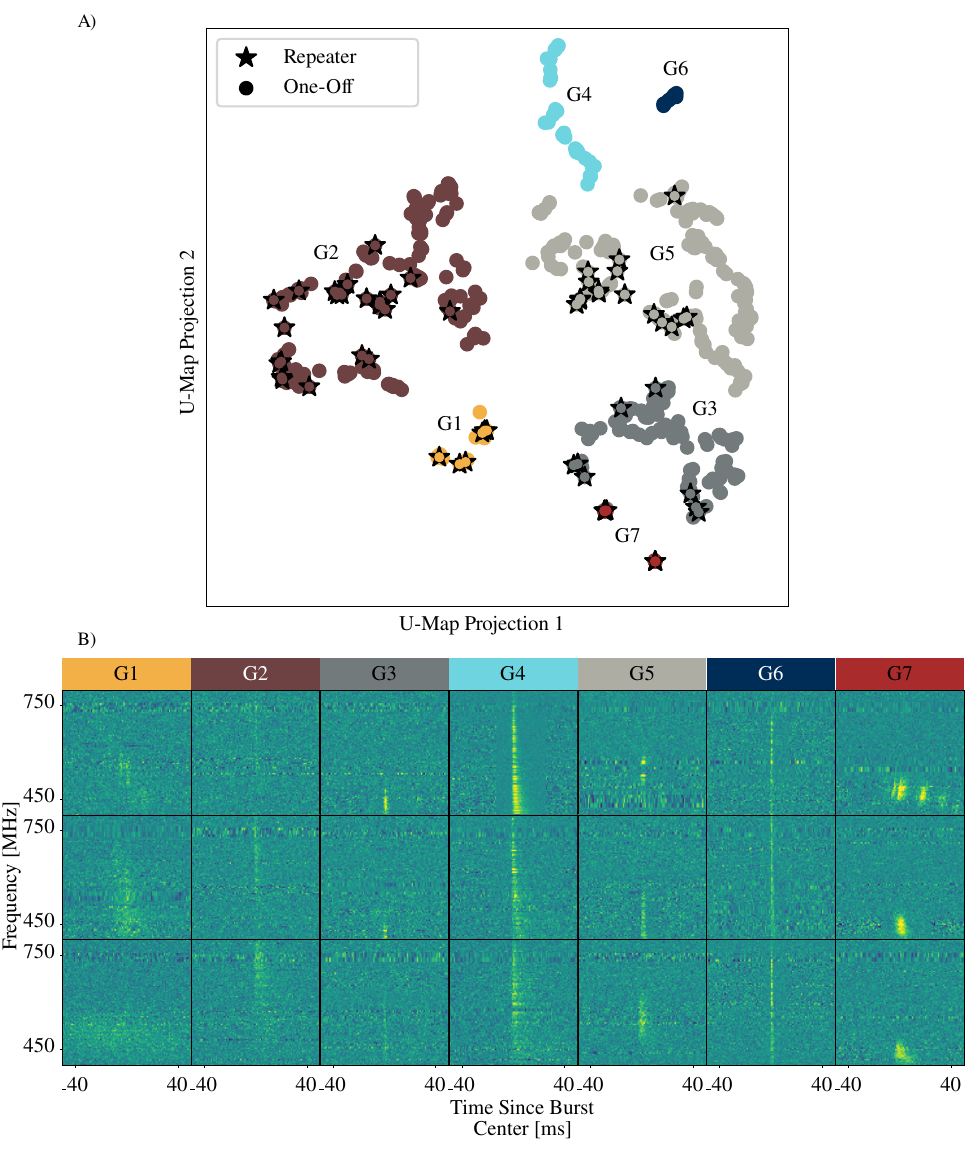}
\end{tabular}
\end{center}
\caption{Unsupervised clustering of the CHIME/FRB DR1 dataset \textit{Top panel:} the UMAP diagram showing the clustering into seven subgroups based on the latent parameters of the CAE model trained on simulations, applied to the DR1 data set. Some clusters (G4, G6, and G7) form distinct, well-separated groups corresponding to previously identified morphologies, while intermediate clusters (G1, G2, G3, G5) show more gradual variation.\textit{Bottom panel:} Three representative bursts from each cluster are shown to illustrate spectral morphology differences. As seen by others \citep[e.g.][]{faber/etal:2024}, some groups exhibit large temporal width but with low frequency dispersion and a `Gaussian-like' morphology (cluster G7). The group also contains bursts with negative linear drifting in central frequency. Other bursts have bandwidth greater than CHIME/FRB's observing band and have fairly uniform flux at each frequency. These bursts can be further divided into those with prominent scattering (e.g. G4) and little to no observable scattering (e.g., G6). \label{fig:types_DR1}}
\end{figure*} 

Taking the 256 latent parameters obtained from CAE embedding, we first conduct principal component analysis (PCA) to map them onto 20 components. PCA is a linear dimensionality reduction method that projects high-dimensional data into a lower-dimensional space while preserving the maximum possible variance \citep{PCA}. We use this preprocessing step to filter out noise, allowing us to focus analysis on more important dimensions during further dimensionality reduction and clustering. For visualization, we then apply UMAP to the standardized ($z$-scored) PCA components, embedding the data into two dimensions. UMAP constructs a graph representation of the high-dimensional data based on nearest-neighbour relationships and then optimizes a low-dimensional embedding that best preserves this original structure \citep{umap}. UMAP is effective at maintaining large-scale relationships between clusters while still resolving fine-grained local structure, making it well suited for visualizing continuous variations in FRB morphology and identifying distinct burst populations. We implement our UMAP using the publicly available \texttt{umap-learn} Python package, with hyper parameters listed in \autoref{table:unsupervised_params}.

\begin{figure*} [htbp!]
\begin{center}
\begin{tabular}{cc}
\includegraphics[width=0.85\textwidth]{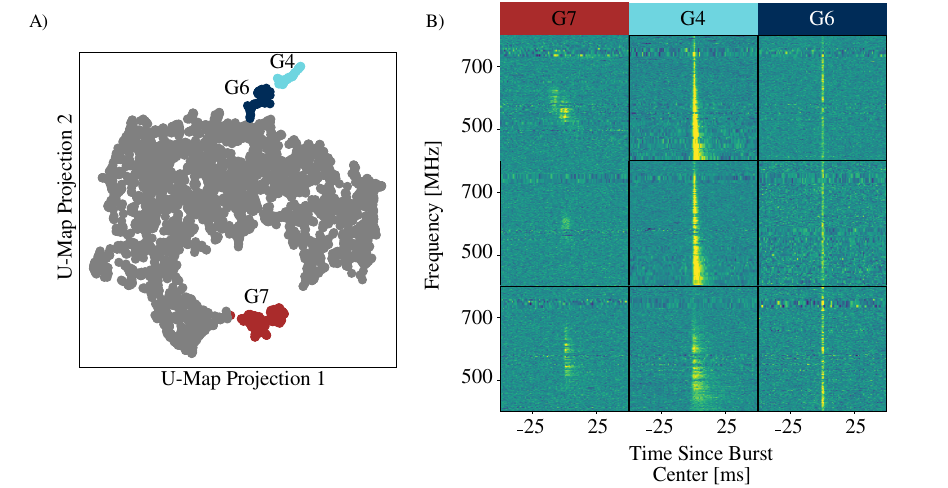}
\end{tabular}
\end{center}
\caption{Unsupervised clustering of the CHIME/FRB DR2 dataset. \textit{Left panel:} the UMAP diagram showing the clustering into seven subgroups based on the latent parameters of the CAE model trained on simulations based on DR1 parameter distributions, applied to the DR2 data set. Clustering of the CHIME/FRB DR2 dataset further enforces our finding that while the majority of FRBs overlap in parameter space, there are three physically distinct groups; Narrowband bursts that may negatively linearly drift (G7), and broadband bursts that are scattered (G4) or not scattered (G6). As more samples are added, intermediate clusters previously seen in \autoref{fig:types_DR1} (G1, G2, G3, G5) merge into one large group. \textit{Right panel:} Three examples of bursts taken from each morphological group identified in both the DR1 and DR2 morphological data to illustrate that they are recovered.\label{fig:types_dr2}}
\end{figure*}
\subsection{HDBSCAN clustering}\label{sec:hdbscan}
We perform clustering on the UMAP latent embeddings using Hierarchical Density-Based Spatial Clustering of Applications with Noise \citep[HDBSCAN;][]{dbs,mcinnes2017hdbscan} as implemented in \citet{scikit-learn}, to look for clusters of similar morphological properties of the FRBs. We set a minimum number of points to form a cluster (min$\_$cluster$\_$size) as 10 points to ensure `large' enough clusters, however further optimization of our clustering procedure is left for future work. This algorithm identifies groups of bursts with similar morphological features by searching for regions of high point density in the reduced parameter space, while naturally handling clusters of varying sizes and densities. Our classification pipeline is shown in \autoref{fig:CAE}. An important advantage of HDBSCAN is its ability to label outliers as noise rather than forcing all data points into clusters, which is particularly valuable when dealing with heterogeneous burst morphologies. We include HDBSCAN parameters in \autoref{table:unsupervised_params}. The resulting clusters form the basis for our morphological classifications which we discuss in \autoref{sec:results}, as we also explore the potential of this approach to discern detectability. 
\section{Results}\label{sec:results}
We apply our classification algorithm trained on DR1-like simulated bursts, as described in \autoref{sec:methods}, to both the DR1 and DR2 datasets. The output of our CAE is both a repeater/one-off burst classification and the latent parameters we reduce in dimensionality and clustered to determine morphological groupings. We first present the unsupervised morphological groupings determined by the CAE approach and connect our findings to existing literature on FRB morphology. We then assess binary repeater classifications generated using mapping from the CAE, and discuss connections between morphological groupings and repeatability.

\subsection{FRB morphological categorization}\label{sec:types}
We find seven subgroups of FRB in CHIME/FRB DR1 data, which later merge to form 4 clearer subgroups in CHIME/FRB DR2 data. While a portion of the FRB population can be separated into distinct clusters using the CAE latent representation, the majority of burst morphologies vary along a continuous gradient (the full distributions of \fitburst\ parameters are given in \autoref{fig:fitburst_params_dr1} in \autoref{sec:appendix}). This distinction between discrete morphological categories, and continuous variation will be used to interpret how morphology relates to repeatability as discussed in \autoref{sec:morph_repeat}. Three groups, identified as G4, G6, and G7 in \autoref{fig:types_DR1} and \autoref{fig:types_dr2}, are clearly distinguishable from other bursts in both DR1 and DR2. Some of the identified clusters match the descriptions of previously qualitatively identified groups (i.e, G7, G6 and G4). Recovering these known morphologies shows that the learned embeddings can capture physical burst properties and independently validates that these visually-defined morphologies are represented in the data. Additionally, our model is able to cleanly separate simple broadband scattered (G4) and unscattered (G6) bursts, suggesting that our latent representation captures additional structure beyond previous manual classification. Some of the groups identified in DR1 (G1, G2, G3 and G5) form a  single continuous distribution when presented DR2, suggesting that these morphologies are part of a broader distribution.

\subsubsection{FRB Subgroups}\label{sec:frb_groups}
Burst morphologies that are clustered separately from the majority of pulses can be split into two broad categories: narrowband bursts that may exhibit downward drifting (G7); and simple broadband bursts (G6 and G4). These simple broadband bursts can further be divided into scattered bursts (G4) and those that show little to no scattering (G6). Understanding how these bursts differ from the broader FRB distribution may yield insight into their origin and environment. Cluster G7 is composed of temporally long and spectrally narrow bursts. Many bursts in the cluster exhibiting multiple sub-bursts that drift linearly downward in central frequency over time (see the top panel in \autoref{fig:types_DR1} and the left panel in \autoref{fig:types_dr2}). This morphology closely matches the ``linear negative drifting repeater" class reported by \citet{faber/etal:2024} and first observed by \citet{Gajjar_2018}. The ability to distinguish G7 in both DR1 and DR2 supports the interpretation that these narrow-band bursts may have a separate emission mechanism or have travelled through a different foreground as compared to the majority of FRBs.

In contrast to G7, clusters G4 and G6 consist of temporally narrow bursts with frequency span greater than or equal to the observing band of CHIME/FRB. These morphological groups can be interpreted as extreme examples of `simple broadband' bursts identified in \citet{pleunis/etal:2021}, representing events that fully occupy the observing band. The distinguishing factor between these groups is that G4 consists of highly scattered bursts, making them appear temporally wider at lower frequencies, while G6 has much lower scattering. This data is consistent with a single population of bursts with an intrinsic broadband, temporally narrow profile. Some bursts from this population travel through some dense plasma scattering screen that causes frequency-dependent temporal broadening as shown in G4 \citep[see e.g.][]{nimmo2025scintillation,pradeep2025scintillation,Jow2026scatter}. Others do not experience this temporal broadening due to the burst propagating through a less homogeneous environment, leading to burst profiles like those in G6. This may hint at differences in the progenitor systems producing G4 and G6 bursts, with G4 burst progenitors residing in/producing less homogeneous environments. 

Although simple broadband bursts have been identified through visual inspection in previous studies \citep{pleunis/etal:2021,faber/etal:2024}, the unsupervised recovery of this morphological group in the CAE latent parameter space indicates that these events can be distinguished through morphological characteristics alone, independent of human annotation.

\subsubsection{Intermediate Property bursts}
\label{sec:intermediate_prop}
In contrast to these visually distinctive classes, the majority of bursts in DR1 and DR2 form a less distinct distribution. Clusters of intermediate property bursts in DR1 are  merged into a broader distribution as more samples are added in DR2.

The three largest DR1 clusters --  G2, G3, and G5 -- form a broader population characterized by Gaussian-like spectral envelopes and comparable temporal widths. sub-bursts in multi-component bursts in these groups possess similar spectral properties. The frequency range for these FRBs range from \(20\%\) to \(>100\%\) of the CHIME/FRB observing bandwidth. These groups are primarily separated by central observing frequency, with G2 containing bursts clustered around 700–800 MHz, G5 containing bursts with central frequency of around 500–600 MHz, and G3 consisting of bursts with central frequencies below 500 MHz (see \autoref{fig:types_DR1}). Visually, these groups are similar to morphological types numbered \#2, \#3 and \#4 as shown in Figure 1 of \citet{kumar/etal:2025}. Aside from these differences in central frequency, the spectral and temporal properties of these groups are largely similar.
\begin{figure}[htbp!]
\begin{center}
\begin{tabular}{cc}
\includegraphics[width=0.30\textwidth]{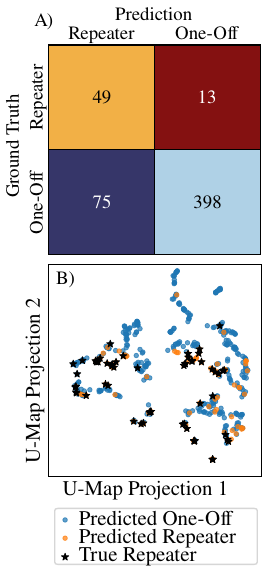}
\end{tabular}
\end{center}
\caption{\textit{Top panel}: Confusion matrices from the CAE classification applied to the CHIME/FRB DR1 data set. 20\% of the true repeaters where classified as one-off bursts, and 16\% of the true one-off bursts were misclassified as repeaters. \textit{Bottom panel}: The UMAP representations of the latent CAE parameters does not show obvious clustering into two distinct groups according to predicted repeatability. Due to the small number of repeaters in DR1, it is difficult to discern any repeatability trends over the embedding space. \label{fig:repeatability_DR1} }
\end{figure}

Cluster G1 seems to be an intermediate case. Bursts in this group have significantly extended temporal durations, more so than G7, and strongly negative spectral running, but are less narrow band and lack instances of downward linear drifting bursts found in G7.  Although separable in DR1, G1 is not distinctly recovered in DR2, where it overlaps with the broader sample (the G2, G3, and G5 groups in \autoref{fig:types_DR1} now form one large group in DR2). This suggests that G1 may correspond to the extremely temporally long bursts in a larger distribution of FRBs, rather than a fundamentally different class.

In DR2, we find that the boundaries between these groups of bursts are not clearly defined, as shown in \autoref{fig:types_dr2}. This suggests that the separation of these groups could reflect a limited sampling of an underlying distribution that these bursts all belong to. The sample size increased in DR2, hence this could indicate that the bursts are not, in fact, distinct groups but rather exist on a continuum of burst morphologies. The disappearance of clear boundaries in DR2 suggests that some apparent clusters in DR1 were likely driven by limited sampling, reinforcing the interpretation that FRB morphology is intrinsically continuous rather than discretely partitioned. Overall, we observe that a significant fraction of the FRB population overlap in morphology space rather than fit into distinct, separable classes. As will be discussed in \autoref{sec:morph_mixed}, this complicates attempts to assign repeatability based on morphology alone.

Observational limitations of CHIME/FRB intensity data could also cause the apparent overlap between repeater and one-off burst populations. Firstly, CHIME/FRB intensity measurements should be interpreted as lower limits of flux in each beam \citep{bridget2023beam}. In addition, the frequency-dependent beam response can distort the observed spectral extent of a burst, causing intrinsically broadband emission to appear artificially narrowband \citep{bridget2023beam}. This effect is illustrated in \citet{sand/etal:2024/morphology}, where the full baseband data of select bursts in the CHIME/FRB intensity catalogue were reanalyzed using more precise beam-forming algorithms. Bursts analyzed this way showed higher bandwidth than those in the CHIME/FRB intensity dataset, indicating that bursts in this dataset are biased towards under reporting bandwidth.

Additionally, because CHIME/FRB is composed of periodically-spaced reflectors, its beam pattern includes not only a main beam but also multiple side lobes. Bursts detected in first order side lobes may have portions of their spectra partially or fully suppressed, causing intrinsically broadband events to appear artificially narrow-band. Approximately 30$\%$ of bursts are estimated to be detected in side lobes, though only bursts detected in higher order side lobes in \citet{dr2} are labelled as originating from a side lobe. Some of the apparent morphological variation among these broadband clusters may therefore reflect observational effects in addition to potential intrinsic differences in burst emission. This effect is important when interpreting how morphology relates to repeatability, as intrinsically broadband (typical one-off) bursts may appear to be narrow-band and therefore closer to repeaters in spectral properties and/or our learned latent representation. Analysis of CHIME catalogue 2 baseband data can provide improved localizations and allow us to separate genuinely narrow-band bursts from instrumental effects.

Finally, FRBs may contain substructure at a shorter duration than the temporal resolution of CHIME/FRB total intensity data as is used in this paper. \citet{faber/etal:2024} identifies that FRBs may have sub-ms sub-burst structure that is not captured in CHIME/FRB intensity data. As discussed in \citet{pandhi2026dispersion,feng2026subburst}, it may be difficult to distinguish downward drifting sub-burst structure from dispersion measure when using a coarser time resolution. As such, downward drifting morphology on short time scales may be interpreted as part of dispersion measure fit and therefore not identified as a distinct morphological feature in CHIME/FRB total intensity data. Future analysis of baseband data from Catalogue 2 can provide a higher temporal resolution, allowing sub-millisecond drifting to be properly resolved.

Taken together, these results indicate that FRB morphology consists of both discrete morphological classes and some continuous structural variation. The persistence of G4, G6, and G7 across data releases suggests physically distinct burst emission properties, while the blending of G1, G2, G3, and G5 in DR2 indicates that some DR1 groupings reflected limited sampling of a broader distribution, rather than distinct subgroups. Having established that the morphologies of CHIME/FRB exhibit both discrete clusters and more continuous variation, we now examine how effectively this morphological parameter space can be used to classify repeatability.

\begin{figure*}[htbp!]
\begin{center}
\begin{tabular}{cc}
\includegraphics[width=0.95\textwidth]{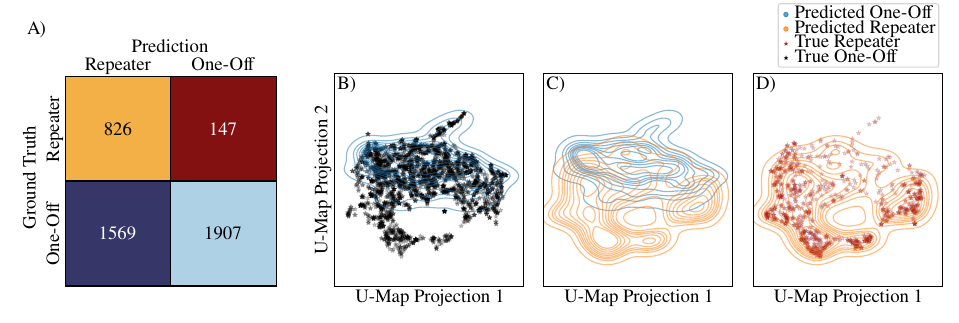}
\end{tabular}
\end{center}
\caption{A) Confusion matrices from the CAE classification applied to the CHIME/FRB DR2 data set. 15\% of the repeaters are misclassified as one-off bursts, while the fraction of one-off bursts misclassified as repeaters grew to 45\% in DR2. The UMAP representations of the latent CAE parameters shown in C) highlights that the latent parameter distribution of predicted repeating and predicted single/one-off bursts overlap significantly for intermediate values. Single one-off bursts are more concentrated towards the top of this distribution, while repeaters are concentrated towards the bottom. These distributions are repeated with the true one-off bursts (B) and true repeaters (D). Morphological categories defined in \autoref{sec:types} can be seen occupying extreme ends of this parameter space. We note that there are a lot of observed singles in the predicted repeater parameter space, as seen by the black points that lie outside the predicted single parameter space in the bottom left panel, resulting in the lower value for the precision and the higher number of misclassified one-off bursts. \label{fig:repeatability_DR2} }
\end{figure*}

\subsection{Repeatability classification performance}\label{sec:repeatability}
Given the central role of repeatability classification in current studies on FRB morphology, we test whether the CAE latent representation trained to encode burst morphology in DR1 can naturally find information about repeatability. As detailed in \autoref{sec:cae}, the CAE trains alongside reconstruction a classifier head that predicts binary classification of repeatability. This allows us to determine whether the features it learns encode information related to repeatability and compare its performance directly to repeatability classifiers \citep[i.e.,][]{arni2025usingdeeplearningrobust,zhang/etal:2020,sun2026dr1_analysis,sun2026dr2_analysis}.

For high-SNR bursts, the CAE achieves performance comparable to parameter-based approaches \citep{zhang/etal:2020,sun2026dr1_analysis,sun2026dr2_analysis}. 
The previous work \citet{sun2026dr2_analysis} finds values for precision of $\mathrm{P}=0.42$ for repeater detection on DR2 while studies of DR1 bursts obtain precision around $\mathrm{P}=0.4$ \citep{Luo_2022unsup,sun2026dr1_analysis}. We obtain: \begin{eqnarray}
   \mathrm{Precision}=0.40&,& \mathrm{Recall}=0.85\,\, \mathrm{(DR1)} \nonumber \\
   \mathrm{Precision}= 0.35&,& \mathrm{Recall}=0.86\,\, \mathrm{(DR2)}. 
\end{eqnarray}
Because apparent one-off bursts may be later shown to repeat (i.e., may they represent the first recorded burst from a repeating source), this measured precision may be lower than the true classification precision. Additionally, broadband bursts detected in CHIME side lobes may appear narrow-band \citep{dr2}. This instrumental effect could contribute up to 30$\%$ of true broadband bursts being observed as narrow-band, and therefore being wrongly classified \citep{dr2}. Taken together, further work is required to determine what proportion of one-off bursts classified to be repeaters are caused by instrumental sampling effects over a true overlap between the morphological distribution between one-off bursts and repeaters. Particularly, CHIME/FRB catalogue 2 baseband data provides improved burst localization, allowing genuinely narrowband bursts to be separated from instrumental artifacts.

Our values of recall are comparable to the recall values of $\mathrm{R}=0.85$ obtained by \citet{kharel/etal:2025} using supervised classification. The high values of recall demonstrate that the CAE can robustly capture repeater characteristics directly from waterfall plots. Although the majority of repeating bursts are correctly identified, a small number of repeaters are categorized as single/one-off bursts by our CAE. As shown in \citet{sun2026dr2_analysis}, repeating sources may produce bursts morphologically more similar to non-repeaters than `typical' repeaters. This suggests that there is diversity of time-frequency structure within repeating bursts, and that repeaters do not always follow the typical repeater morphology discussed in \citet{pleunis/etal:2021}. A detailed discussion of repeating bursts with non-repeater-like morphology is presented in \autoref{sec:single_morph}. Overall, the CAE-based classifier achieves repeatability prediction performance comparable to methods based on parametric burst properties, despite operating directly on waterfall plots. The high recall indicates that the latent representation captures morphological features commonly associated with repeating bursts. However, the presence of repeaters that are classified as one-off bursts reflects the known diversity of repeater morphologies, suggesting that repeating sources do not always produce bursts with the canonical time–frequency structure typically associated with repeaters. This motivates a closer examination of how repeatability is distributed across the morphological clusters identified in the CAE latent space.

\subsection{Are morphology and repeatability linked?}\label{sec:morph_repeat}
Repeating bursts do not correspond to a single morphological class, and repeaters are not uniformly distributed across the clusters identified by the CAE. Instead, some clusters represent extreme ends of the morphology spectrum associated with repeating and non-repeating sources. Clusters G4 and G6 (as shown in \autoref{fig:repeatability_DR2}) are dominated by one-off burst events and exhibit morphologies characteristic of non-repeaters, including very broad bandwidth, and power-law negative drifting. In contrast, G7 contains a high fraction of repeating bursts and is characterized by short-bandwidth, long-temporal-width bursts typical of repeaters. These trends suggest that while morphology alone does not determine repeatability, certain morphological groups are preferentially associated with repeaters or non-repeaters.
\begin{figure*}[htbp!]
\begin{center}
\begin{tabular}{c}
\includegraphics[width=0.95\textwidth]{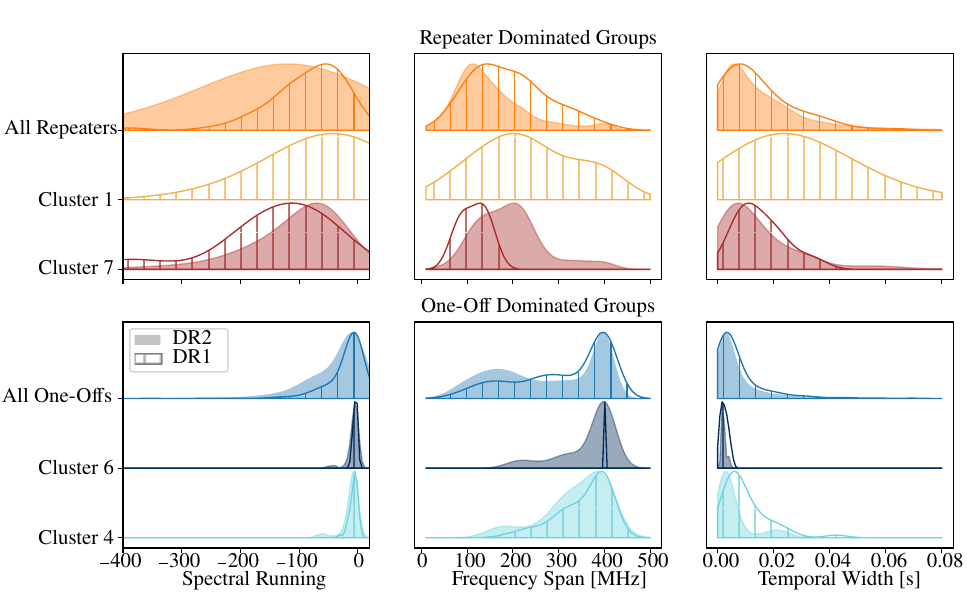}
\end{tabular}
\end{center}
\caption{Parameter distributions from the \fitburst\ model for the morphological groupings described in \autoref{sec:types} obtained when applying our CAE unsupervised clustering to the CHIME/FRB DR1 as compared to the full one-off and repeater distributions of CHIME/FRB DR1 and DR2. Spectral running, temporal width, and frequency span are shown as these have been established to be most distinguishing between repeater and one-offs. We find that repeater-rich clusters occupy the same region of parameter space as the global repeater population, despite repeatability not being used in the clustering procedure. Conversely, clusters with no repeaters exhibit even lower spectral running, shorter temporal durations, and broader bandwidths than the average non-repeating burst. The full set of \fitburst\ parameters for DR1 is shown in \autoref{fig:fitburst_params_dr1}. \label{fig:fitburst_params_repeat}}
\end{figure*}
To quantify the differences in morphology between repeaters and one-off bursts, we examine the distribution of \fitburst~model parameters within each CAE grouping in \autoref{fig:fitburst_params_repeat}, and compare them based on previously established work on repeater/one-off burst morphology. Previous studies show that temporally-broad and spectrally-narrow bursts with more negative spectral running are associated with repeaters \citep{pleunis/etal:2021}. Particularly, highly negative spectral running has been shown to be the strongest individual predictor of repeatability \citep{sun2026dr1_analysis}. 

\subsubsection{One-off-Dominated Clusters}\label{sec:single_morph}
Clusters G4 and G6 define morphologies that are overwhelmingly associated with one-off sources in both DR1 and DR2 (shown in \autoref{fig:types_DR1} and \autoref{fig:types_dr2}). Bursts in these groups exhibit large frequency spans (as shown in \autoref{fig:fitburst_params_repeat}), often spanning nearly the full CHIME/FRB observing band. The persistence of these clusters suggests that broadband (bandwidth $>$ 400 MHz) emission is a common morphological feature among one-off FRBs. Although both G4 and G6 are dominated by broadband emission, they differ in scattering, suggesting that they may originate from or propagate through different environments.

G6 and G4 can again be interpreted as an extreme subset of the `simple broadband' bursts, which were associated primarily with non-repeating sources \citep{pleunis/etal:2021}. Compared to repeater-dominated clusters, bursts in G4 and G6 exhibit relatively small spectral running values, consistent with \citet{pleunis/etal:2021,kharel/etal:2025,sun2026dr2_analysis}, who find that low spectral running is a predictor of non-repeating bursts. The corresponding \fitburst\ parameter distributions for clusters G4 and G6 as compared to the distributions for all non-repeaters can be seen in the bottom panel of \autoref{fig:fitburst_params_repeat} for DR1 (hashed distribution) and DR2 (solid curves).

We used a Student's $t$-test to determine whether the parameter distributions of the G4 and G6 clusters differ significantly from the parameter distributions of remaining non-repeater population, we give these values in parentheses below.

Cluster G6 in particular shows significantly different parameters than the broader non-repeater population; the mean spectral running of G6 bursts is $-7.50$, compared to $-47.1$ for the broader one-off population ($t$-test value $p = 0.005$), the mean temporal width is $2.2$ ms versus $8.6$ ms for all one-off bursts ($t$-test value $p = 0.0003$), and the mean frequency span is $356$ MHz compared to $275$ MHz ($t$-test value $p = 3\times10^{-10}$, as shown in \autoref{fig:fitburst_params_repeat}). These results demonstrate that G6 may represent a subpopulation of simple broadband non-repeater bursts with comparatively clean propagation paths, with minimal plasma lensing and scattering.

\begin{figure}[htbp!]
\begin{center}
\begin{tabular}{c}
\includegraphics[width=0.45\textwidth]{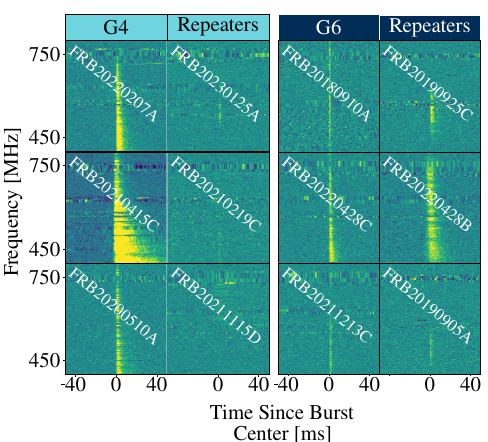}
\end{tabular}
\end{center}
\caption{Examples of repeating bursts in DR2 found in G4 and G6, clusters comprised predominantly of one-off bursts. An additional burst from these repeating sources not considered part of these primarily one-off sources are also shown to the left, and displays morphology more typical of repeaters.  Despite being repeaters, these bursts show broadband and, in the case of G4, negative power-law drifting burst structure commonly associated with one-off bursts. \label{fig:dr2_exceptions}}
\end{figure}

The overwhelming majority of bursts in G4 and G6 originate from non-repeaters. No repeaters are present in these clusters in DR1, and only a small number appear in DR2 (4 for G4 and 3 for G6). The waterfall plots of repeaters in these groups are visually indistinguishable from the predominantly one-off population (see \autoref{fig:dr2_exceptions} for example waterfall plots). This indicates that the morphologies captured by G4 and G6 are not unique to one-off events, although they remain strongly associated with non-repeaters. The physical origin of this morphology remains uncertain, however we discuss the physical interpretation of repeater-dominated groups below in \autoref{sec:multi_morph}. 

\subsubsection{Repeater-dominated morphological groups}\label{sec:multi_morph}
Extreme examples of morphology previously associated with repeater bursts \citep{pleunis/etal:2021,curtin/etal:2025} can be seen in cluster G7. This cluster contains bursts with significantly larger temporal widths (mean $= 15.4$ ms) than other clusters, as well as systematically narrower frequency spans (mean $= 194$ MHz) and more negative spectral running (mean $= -113$) when compared to other groups in DR2. This is consistent with previous CHIME/FRB analyses which show that repeaters tend to be longer in duration, narrower in bandwidth, and exhibit stronger downward spectral evolution compared to one-off bursts \citep{pleunis/etal:2021,chime/2022}. These trends are illustrated in the bottom panel of \autoref{fig:fitburst_params_repeat}, which compares the distributions of temporal width, frequency span, and spectral running for G7 to all repeaters in DR2, and other identified clusters (we note that cluster G1 does not occur in DR2). Qualitatively, many bursts in G7 display the linear, downward-drifting sub-burst structure identified in baseband analyses by \citet{sand/etal:2024/morphology}, which was shown to be characteristic of repeating FRBs. Consistent with this interpretation, in DR1 and DR2, G7 has the highest repeater fractions across all morphological clusters. The recovery of this morphology through unsupervised clustering indicates that the characteristic spectral and temporal properties of a subgroup of repeating FRBs occupy a distinct region of the learned morphology space.  One possible way to interpret the spectral properties of these bursts is through the relativistic shock model proposed by \citet{Metzger_2022}:
\begin{equation} 
n\propto r^k;  k \in [-1,1],
\end{equation}
Within this framework, the density profile could result from ejecta from prior bursts, and these events could correspond to emission with minimal intrinsic frequency drifts or to shocks with central frequency outside of the CHIME/FRB observing band. This scenario is consistent with sources that have emitted multiple bursts, since these bursts could produce ejecta to enrich the upstream medium, and provides a potential physical explanation for the repeated activity and spectral properties observed in G7.
\subsubsection{Mixed-morphology clusters}\label{sec:morph_mixed}
Most bursts in CHIME/FRB DR2 are characterized by our method as belonging to a large cluster (shown in grey in \autoref{fig:types_dr2}) contain a mixture of repeating and one-off bursts, indicating that there remains morphological overlap between repeaters and one-off events. This overlap may result from the presence of non-repeating bursts with repeater-like time–frequency structure, one-off bursts originating from repeaters being misclassified due to incomplete sampling, intrinsic overlap in the physical mechanisms producing FRBs, and instrumental effects artificially lowering the observed bandwidth of broad-band bursts. Within this mixed group, repeaters tend towards significantly narrower bandwidths than one-off bursts. In contrast, the mean temporal width and spectral running do not differ significantly between repeating and non-repeating bursts in this cluster. This mixed cluster provides evidence that morphology alone does not provide a definitive indicator of repeatability, even when considering regions of the parameter space rich in repeating/non-repeating bursts.
\begin{figure*}[htbp!]
\begin{center}
\begin{tabular}{c}
\includegraphics[width=0.7\textwidth]{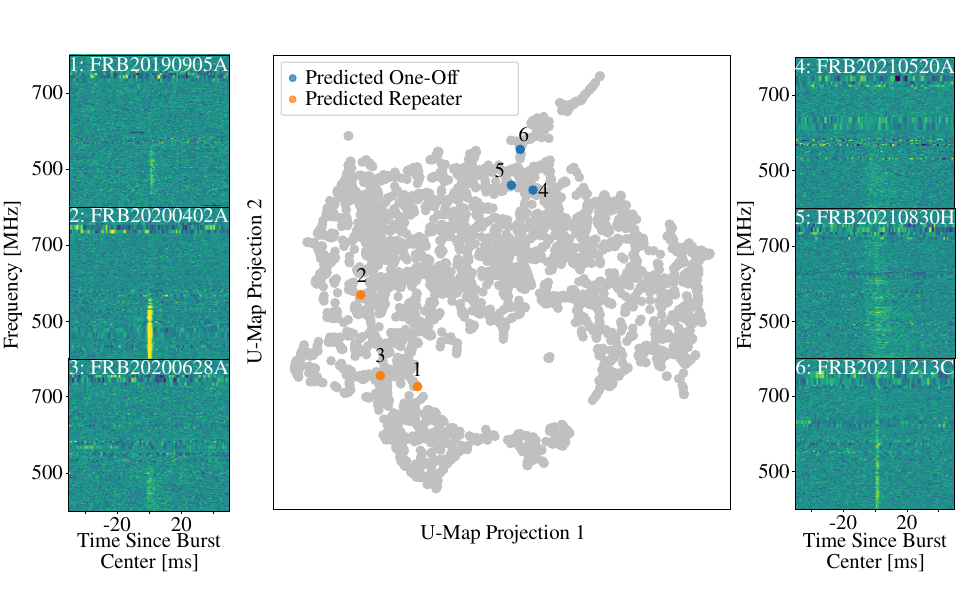}
\end{tabular}
\end{center}
\caption{Individual bursts observed from the repeating burst FRB 20190905A. Bursts are numbered in order of arrival time, with 1 being the oldest recorded burst. Predictions from our classification pipeline are indicated by the colour of the point, with orange denoting that it was predicted to be a repeater, while blue denotes that the observation was predicted to be a one-off burst from its morphological clustering. Under our spectral morphology classification, the three earliest bursts are correctly identified as being generated by a repeating source (orange dots) while the later three are misclassified as one-off bursts (blue dots). Looking at their frequency-time plots shown to the left and right of the plot, earlier bursts exhibit the more typical narrow-band structure with a temporally-long morphology, while later bursts have a broad-band spectral signature. This is especially the case with the most recent burst from this source, which exhibit the very temporally short and broadband morphology characteristic of cluster G6, a cluster with burst morphology typical of one-off bursts. \label{fig:20190905a}}
\end{figure*}

\subsubsection{Repeaters with Outlying Morphology}
We also detect repeating bursts occupying regions of parameter space primarily associated with non-repeaters, demonstrating that broadband morphologies are not exclusive to apparently one-off bursts. Notably, \autoref{fig:20190905a} shows that even within a single repeating source, burst morphology can vary dramatically between typically repeater bursts or typical single/one-off bursts. In the case of FRB 20190905A, early bursts exhibit narrow bandwidths and steep spectral running, while later bursts are significantly more broadband with weaker spectral running, closely resembling one-off events. Morphology not only varies between populations of bursts, but also between repeated bursts emitted from the same source. This intra-source diversity further supports our finding that morphology alone cannot be used to separate repeating and non-repeating bursts; the overlap between repeater and non-repeater populations may be caused in part by intrinsic variability in emission processes.

\subsubsection{Summary and Implications}
While repeaters tend to be temporally longer and shorter in bandwidth than one-off bursts, the presence of both repeaters and one-off bursts within several morphological clusters re-enforces our interpretation in \autoref{sec:morph_mixed} and \autoref{sec:single_morph} that we cannot currently sharply define boundary between repeating and non-repeating FRBs based on their morphology alone. This overlap may arise from differences in the surrounding environments in which bursts are emitted \citep{Metzger_2022}, intrinsic variation in burst emission mechanisms, instrumental limitations due to broad-band bursts appearing artificially narrow-band due to instrumental effects, or the possibility that some apparently one-off bursts originate from repeaters that have not yet been observed to repeat.

In their analysis of microsecond-resolution raw voltage data, \citet{curtin/etal:2025} also finds that repeaters tend to have longer duration and shorter band-width than one-off bursts. However, after normalizing for sub-burst duration and separation time by the total duration of each burst, they find that the distributions of the parameters are statistically consistent, suggesting that repeaters and one-off FRBs could share a common emission mechanism. Through analyzing prolifically repeating sources, recent studies find that the brightest bursts from repeaters can resemble non-repeaters, suggesting that some apparently non-repeating FRBs may represent the high-energy end of the repeater population rather than a distinct class of event \citep{kirster2024rep,ould2026rep}.

Together, these results suggest that morphology provides statistical information about repeatability, rather than serving as a definitive indicator of whether a source repeats. Further analysis on which proportion of these intermediate morphologies are caused by overlap in intrinsic burst mechanism or instrumental effects is needed to determine the extent to which the observed continuum of FRB morphology reflects intrinsic FRB morphological diversity.

\section{Conclusion}\label{sec:end}
We presented an unsupervised classifier that can identify groups of FRBs based on their spectral morphology. While previous groups define morphological groups based on visual inspection of waterfall plots, our methodology is reproducible and data-driven, using automatic feature extraction and clustering to define morphology objectively. While some of our groupings mirror those presented in the literature, our classifier reveals additional FRB groupings that were not previously identified through manual inspection. We find that:
\begin{itemize}
    \item{} Burst morphology derived using current CHIME/FRB waterfall plots alone does not provide a clear distinction between repeating and non-repeating bursts. Although morphology can probabilistically predict repeatability, there is significant morphological overlap between repeaters and single/one-off bursts.
    \item{} Additionally, some more extreme morphology groups can be separated from the bulk of observed FRBs. Using a CAE, we are able to distinguish morphological groups that correspond to those previously identified in the literature \citep[e.g.][]{kumar/etal:2025}.
    \item{} Repeaters are sometimes associated with narrow-band, long duration bursts that exhibit negative spectral running (\autoref{fig:fitburst_params_repeat}), but these morphological parameters vary along a continuum for the objects presented in CHIME/FRB DR1 and DR2, and therefore do not present a clearly-defined class. Some of these non-repeaters with repeater-like properties could arise due to instrumental effects. Particularly, non-repeaters detected in CHIME side lobes can appear more narrow-band and therefore similar to repeaters. Further work is required to distinguish side-lobe bursts from genuinely narrow-band bursts.
     \item{} We find two distinct clusters of bursts primarily dominated by single-detection bursts, differing only by the fact that one contains scattered bursts while the other does not. This suggests that they could both originate from the same primarily one-off burst population, and differ only by the presence of scattering due to plasma screens between their source and the observer.
    \item{} While some `obvious' or extreme morphological groups exist that are primarily populated by repeater or non-repeater bursts, as shown in \autoref{fig:types_dr2}, a large number of repeaters and one-off burst FRBs share similar properties in morphological parameter space.
    \item{} Analysis using current CHIME/FRB total intensity data is limited by the fact that sub-burst structure below the CHIME/FRB intensity temporal resolution cannot be determined, and that bursts in the grating lobes may have part of their spectra suppressed. Using CHIME/FRB catalogue 2 baseband data for similar analysis will account for these limitations.
\end{itemize}

In addition to unsupervised morphological categorization, our approach uses a semi-supervised approach to distinguish repeaters from apparent non-repeating bursts. We find that our approach performs comparably with methods that classify repeatability based on parameters derived of parametric fits to FRB waterfall plots. While we find no clear association between the location of the burst in our derived morphology space and its repeatability, different morphological categories exhibit differing tendencies towards repeaters or non-repeaters. This includes clusters G4 and G6, which are primarily dominated by single-detection bursts, and cluster G7, which matches the morphology for downward linear drifting repeaters described in \citet{sand/etal:2024/morphology}.

\section{Acknowledgements}
We thank Ketan Sand and Ayush Pandhi for useful comments on our manuscript, and Gwen Eadie and Bryan Gaensler for discussions on the ideas and methods.
We thank the Digital Research Alliance of Canada for compute resources. Funding was provided by grants from the Natural Sciences and Engineering Research Council of Canada (RGPIN-2025-06483 and SMFSU-60768) the Connaught Fund. The Dunlap Institute is funded through an endowment established by the David Dunlap family and the University of Toronto. 
The authors at the University of Toronto acknowledge that the land on which the University of Toronto is built is the traditional territory of the Wendat Nation, the Seneca, and the Mississaugas of the Credit. Today, this meeting place is still the home to many Indigenous people from across Turtle Island. They are grateful to have the opportunity to work on this land.
\bibliography{frbmorph} 
\section{appendix}
\label{sec:appendix}
The following tables include the range of parameters used for both the supervised and unsupervised classification methods described in \autoref{sec:methods}.

\begin{table}[htbp!]
\caption{Range of \fitburst~parameters used to generate simulated repeater and one-off bursts. \label{table:fitburst_params}} 
\begin{center}    
\begin{tabular}{ |c|c|c|c|}
\hline
\rule[-1ex]{0pt}{3.5ex} Parameter  & Repeater Central Value & one-off burst Central Value & Range\\ 
\hline
\rule[-1ex]{0pt}{3.5ex} burst$\_$width &  0.00983 & 0.00393 & [0.126,0.001]\\
\hline
\rule[-1ex]{0pt}{3.5ex} ref$\_$freq & 400.1953125 MHz & 400.1953125 MHz & --\\
\hline
\rule[-1ex]{0pt}{3.5ex} scat$\_$time & 0.0022 & 0.0012 MHz & [0.09,0.0001]\\
\hline
\rule[-1ex]{0pt}{3.5ex} spec$\_$ind & 46.65 & 3.3 & [99,-9.4]\\
\hline
\rule[-1ex]{0pt}{3.5ex} spec$\_$run & -69.7 & -8.5 & [10.3,-910]\\
\hline
\end{tabular}
\end{center}
\end{table}
\begin{table}[htbp!]
\caption{Unsupervised learning hyper-parameters for the CAE applied to the CHIME/FRB DR1 and DR2 data. \label{table:unsupervised_params}} 
\begin{center}    
\begin{tabular}{ |c|c|c| }
\hline
\rule[-1ex]{0pt}{3.5ex} Parameter & Algorithm  & value\\ 
\hline\hline
\rule[-1ex]{0pt}{3.5ex} n$\_$neighbours & UMAP & 5\\
\hline
\rule[-1ex]{0pt}{3.5ex} n$\_$components & UMAP & 2\\
\hline
\rule[-1ex]{0pt}{3.5ex} min$\_$dist & UMAP & 0.01\\
\hline
\rule[-1ex]{0pt}{3.5ex} min$\_$cluster$\_$size & HDBscan & 20\\
\hline
\end{tabular}
\end{center}
\end{table}
\begin{figure*}[htbp!]
\begin{center}
\begin{tabular}{c}
\includegraphics[width=0.85\textwidth]{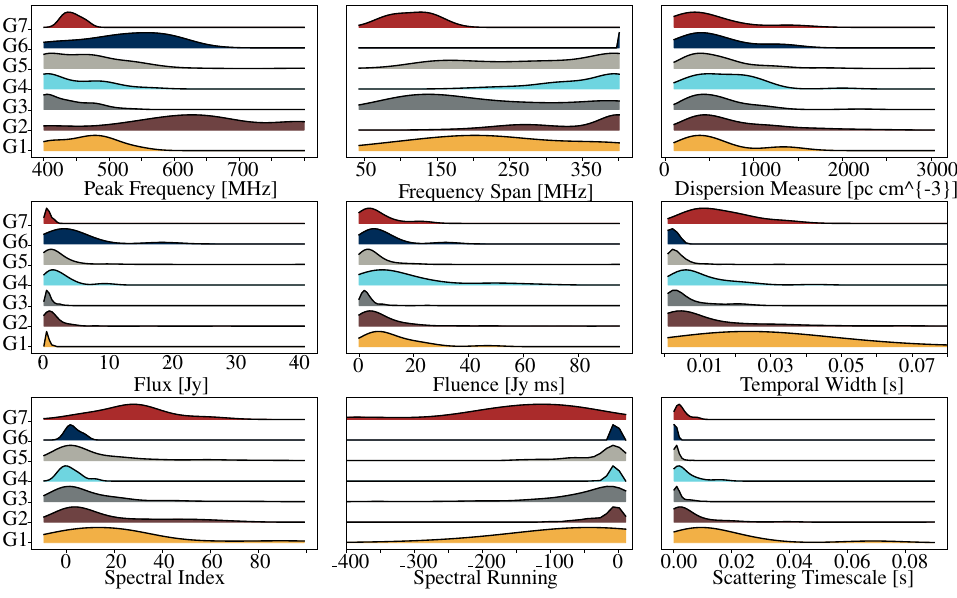}
\end{tabular}
\end{center}
\caption{The full suite of parameter distributions from the \fitburst\ model for the morphological groupings described in \autoref{sec:types} obtained when applying our CAE unsupervised clustering to the CHIME/FRB DR1. A subset of parameter distributions relevant to repeatability are shown in \autoref{fig:fitburst_params_repeat}.\label{fig:fitburst_params_dr1}}
\end{figure*}
\begin{figure*}[htbp!]
\begin{center}
\begin{tabular}{c}
\includegraphics[width=0.85\textwidth]{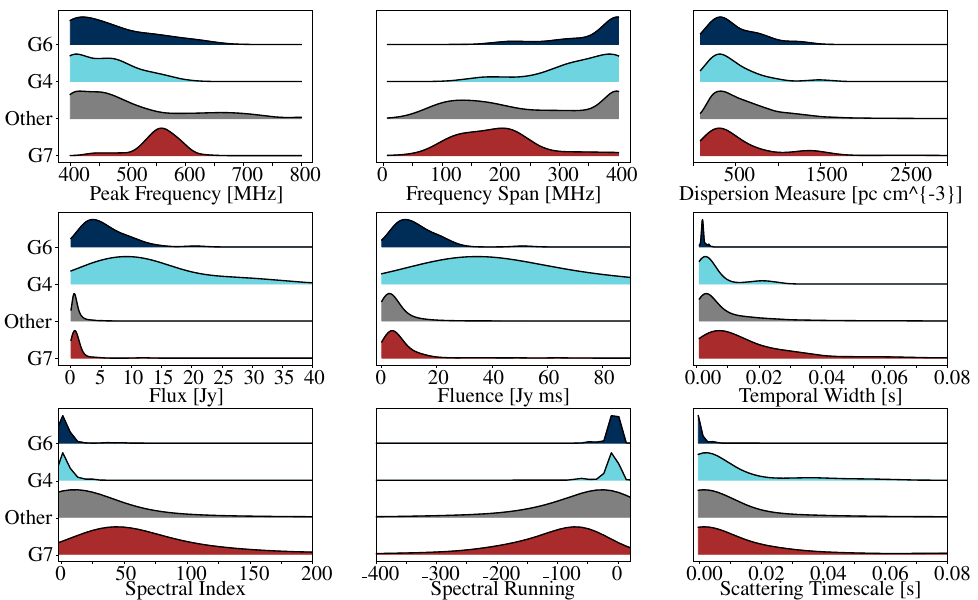}
\end{tabular}
\end{center}
\caption{Parameter distributions from the \fitburst\ model for the morphological groupings described in \autoref{sec:morph_repeat} obtained when applying our CAE unsupervised clustering to the CHIME/FRB DR2. A subset of parameter distributions relevant to repeatability are shown in \autoref{fig:fitburst_params_repeat}, where the bottom panel indicates the distribution of \fitburst\ parameters for DR2 repeaters. \label{fig:fitburst_params_dr2} }
\end{figure*}

\end{document}